\documentclass[twocolumn]{aastex7}

\usepackage[version=4]{mhchem}

\shorttitle{Sub-Neptune magma clouds}
\shortauthors{Lee, Werlen \& Dorn}
\received{---}
\revised{---}
\accepted{---}

\submitjournal{ApJL}

\begin{document}

\title{Mineral cloud formation above magma oceans in sub-Neptune atmospheres}

\author[orcid=0000-0002-3052-7116,gname=Elspeth,sname=Lee]{Elspeth K. H. Lee}
\affiliation{Center for Space and Habitability, University of Bern, Gesellschaftsstrasse 6, CH-3012 Bern, Switzerland} 
\email{elspeth.lee@unibe.ch}

\author[orcid=0009-0005-1133-7586,gname=Aaron,sname=Werlen]{Aaron Werlen}
\affiliation{Institute for Particle Physics and Astrophysics, ETH Zurich, CH-8093 Zurich, Switzerland} 
\email{awerlen@student.ethz.ch}

\author[orcid=0000-0001-6110-4610,gname=Caroline,sname=Dorn]{Caroline Dorn}
\affiliation{Institute for Particle Physics and Astrophysics, ETH Zurich, CH-8093 Zurich, Switzerland} 
\email{dornc@phys.ethz.ch}

\begin{abstract}
The potential presence of a magma surface below a thick atmosphere primarily composed of hydrogen in some sub-Neptune exoplanets suggests a strong link between the interior composition and atmosphere through chemical coupling of volatile and refractory species. 
In this study, we aim to model the possibility for mineral cloud formation in the atmosphere of sub-Neptunes from outgassing of refractory species at the magma surface.
In our specific cases, we find that mineral clouds easily form near the magma-atmosphere boundary, but also higher in the atmosphere once vapour is mixed to the cooler atmospheric regions.
We find that the vertical cloud structure depends on the mixing profile of the atmosphere, with stronger mixing allowing particles to remain lofted in the atmosphere, while weak to moderate mixing produces larger, more sedimented cloud particle profiles.
We suggest that due to the strong thermal feedback from cloud opacity, clouds may play an important role in the overall structure of the interior-surface-atmosphere coupled system in sub-Neptunes, as well as affect their observed spectral properties, especially at near-infrared wavelengths.
\end{abstract}


\section{Introduction}

With the recent possibility to characterize sub-Neptune atmospheres to great detail with the James Webb Space Telescope (JWST) \citep[e.g.][]{madhusudhan_carbon-bearing_2023,holmberg_possible_2024,benneke_jwst_2024,schmidt_comprehensive_2025,felix_evidence_2025}, a concerted effort in the field has been undertaken to understand the complex coupled system of atmosphere, surface and interior, their chemical compositions, temperature-pressure profiles, evolution through time and dependence on bulk planetary properties \citep[e.g.][]{ginzburg_super-earth_2016, chachan_role_2018, kite_superabundance_2019, schlichting_chemical_2022, kite_water_2021, dorn2021hidden, luo_interior_2024, tian_atmospheric_2024, seo_role_2024}.

The majority of observed sub-Neptunes may host magma oceans as the planets form hot due to their gravitational binding energy being converted to heat, and their long (Gyrs) cooling timescales.
Temperatures below the hydrogen-dominated envelopes after formation can reach $10^4$~K, which is an upper limit for a stable H/He envelope \citep{ginzburg_super-earth_2016}. 
Thick hydrogen-dominated envelopes act as an insulating layers as \ce{H2}-\ce{H2} collisional induced opacity (CIA) is a strong greenhouse gas at atmospheric pressures $\gtrsim$ 10 bar.
Due to this opacity, large \ce{H2} atmospheres imply long term cooling timescales of such planets \citep[e.g.][]{Lopez2014}, which allows magma oceans to exist over gigayear timescales. 

 The boundary between the magma ocean and the atmosphere is compositionally coupled, chemically reactive, and thermally active \citep{schaefer2018magma,kite_atmosphere_2020}. 
 For example, redox reactions between primordial hydrogen and magma produce water and other volatiles with significant consequences on the mean molecular weight of the atmosphere \citep{benneke_jwst_2024}. 
 The fact that chemical coupling at the atmosphere-magma ocean boundary exists allows us to learn from spectroscopically retrieved atmospheric composition about the deeper interior. 
 Although these links are not direct, they involve models of atmospheric structure, photochemistry, as well as chemical equilibration \citep{Werlen2025}. 
 The majority of exoplanet interior models do not include this complexity which hampers the interpretation of atmospheric metallicity and composition to accurately infer bulk composition.

 An additional complexity in interpreting exoplanet atmospheres arises due to the possible presence of clouds and haze \citep[e.g.,][]{benneke2012atmospheric,line2016influence}. 
 Before data of JWST were available for sub-Neptunes, spectral features were expected for hotter sub-Neptunes and longer wavelengths ($>$2-3 $\mu m$) \citep{crossfield2017trends,benneke2019sub}. 
 Recent observations do show some clear features in temperate sub-Neptunes \citep[e.g.][]{madhusudhan_carbon-bearing_2023,holmberg_possible_2024,benneke_jwst_2024}, while this is less evident for hotter sub-Neptunes \citep[e.g.][]{Brande2024}. 
 High-altitude clouds may significantly limit atmospheric characterization by muting spectral features. 
 Clouds add degeneracy in atmospheric retrieval and limit constraints on atmospheric metallicities \citep{welbanks2025challenges}.
 A profound understanding of cloud formation in sub-Neptunes with coupled atmosphere-interiors is limited and yet it is a central aspect in the interpretation of spectral data from missions like JWST as well as ELT and ARIEL.

In this letter, we investigate the potential for mineral cloud formation in the atmosphere of sub-Neptunes containing a magma ocean surface, driven by the outgassing of refractory material from the magma ocean into the atmosphere.
In Section \ref{sec:theory}, we detail the microphysical cloud formation model and application to the self-consistently produced interior-surface-atmosphere sub-Neptune models produced by \citet{Werlen2025}.
In Section \ref{sec:results}, we show the results of the cloud formation model for various vertical mixing profile scenarios, as well as produce synthetic transmission and emission spectra of the models to elucidate the observable consequences of cloud formation. 
Section \ref{sec:discon} contains the discussion and conclusions of our study.

\section{Cloud formation above magma oceans}
\label{sec:theory}

\begin{table}[]
    \centering
    \caption{Volume mixing ratio (VMR) of gaseous species outgassed from the magma surface as modelled in \citet{Werlen2025}.}
    \begin{tabular}{c|c} \hline \hline
        Species & VMR \\ \hline 
        \ce{H2}  & 0.997 \\
        CO  & 1.501$\cdot$10$^{-8}$ \\
        \ce{CO2}  & 5.080$\cdot$10$^{-12}$ \\
        \ce{CH4}  & 2.520$\cdot$10$^{-4}$ \\
        \ce{O2}  & 4.234$\cdot$10$^{-14}$ \\
        \ce{H2O}  & 2.453$\cdot$10$^{-3}$ \\
        Fe  & 2.187$\cdot$10$^{-5}$\\
        Mg  & 7.650$\cdot$10$^{-5}$ \\
        SiO  & 2.861$\cdot$10$^{-4}$ \\
        Na  & 5.454$\cdot$10$^{-5}$ \\ \hline
    \end{tabular}
    \label{tab:surf_VMR}
\end{table}

\begin{table*}[]
    \centering
    \caption{Properties of the cloud formation species considered in this study.}
    \begin{tabular}{l|l|l|l} \hline \hline
        Species & Formation reaction & Vapour pressure & Optical constants \\ \hline 
        \ce{SiO2}  & SiO + \ce{H2O} $\rightarrow$ \ce{SiO2}[s] + \ce{H2} & \citet{Woitke2018} & \citet{Palik1985, Zeidler2013} \\
        \ce{MgSiO3}  & Mg + SiO + 2\ce{H2O} $\rightarrow$ \ce{MgSiO3}[s] + 2\ce{H2} & \citet{Visscher2010} & \citet{Dorschner1995} \\
        \ce{Fe}  & Fe $\rightarrow$ Fe[s] & \citet{Visscher2010} & Lynch \& Hunter in \citet[][]{Palik1991} \\
        \ce{Na2S}  & 2Na + \ce{H2S} $\rightarrow$ \ce{Na2S}[s] + \ce{H2} & \citet{Morley2012} & \citet{Montaner1979,Khachai2009}\\ \hline
    \end{tabular}
    \label{tab:cond_sp}
\end{table*}

For the cloud microphysical model, we use the equation set described in \citet{Lee2025} coupled to a 1D vertical advection and diffusion scheme \citep{Lee2025b} suitable for modeling the vertical transport of cloud particles and condensable vapor in an atmosphere.
The \citet{Lee2025} microphysical cloud formation model uses the mass moment method to evolve the cloud properties assuming a monodisperse size distribution. 
This scheme includes nucleation, condensation, evaporation and collisional growth processes of cloud particles in the atmosphere. 
We augment the \citet{Lee2025} model to include the ability to evolve mixed material grains in a similar manner to the \citet{Helling2008} methodology, through including the time evolution of an additional first mass moment for each cloud formation species. 

As input to the cloud microphysical model, we use the coupled interior-surface-atmosphere, self-consistently produced temperature-pressure (T-p) and chemical profiles from \citet{Werlen2025}, taking their $T_{\rm surf}$ = 3000 K magma-atmosphere boundary, 6 M$_{\oplus}$, 2.5 R$_{\oplus}$ (log g $\approx$ 3) model as input to the cloud formation scheme. 
The profiles were produced by first chemically equilibrating the global composition of a sub-Neptune to determine the atmosphere in equilibrium with an underlying magma ocean. 
Details of the chemical equilibrium thermodynamic modeling are discussed in \citet{Werlen2025} and \cite{schlichting_chemical_2022}. 
Second, the output of the equilibrium mixing ratios and the surface pressure-temperature condition (3000 K, 10$^5$ bars) were passed further to self-consistently calculate the mixing ratios and the T-p profile throughout the atmosphere up to the upper observable atmosphere.
To this end, \citet{Werlen2025} first used FastChem \citep{stock_fastchem_2018}, a chemical equilibrium code that computes atmospheric compositions, HELIOS-K \citep{grimm_helios-k_2021}, a line-by-line opacity calculator, and HELIOS \citep{malik_helios_2017, malik_self-luminous_2019}, a radiative–convective equilibrium solver, to compute an initial T–p profile.
This profile, along with the mixing ratios from the chemical equilibrium model, was passed to the 1D photochemical model VULCAN \citep{Tsai2021}, which accounts for disequilibrium chemistry driven by vertical mixing and photochemistry.
To capture the feedback of photochemistry on the thermal structure, the updated abundances from VULCAN were then passed back to HELIOS-K and HELIOS. This iterative process was repeated until convergence was reached.

The mixing ratios calculated from the chemical equilibrium model in \cite{Werlen2025} are presented in Table \ref{tab:surf_VMR}, which are also taken as the surface boundary values for the cloud microphysical model.
From this selection of outgassed species available from \citet{Werlen2025}, we can construct a list of potential mineral cloud formation species and their formation reactions in the atmosphere informed from the gas phase composition.
This is presented in Table \ref{tab:cond_sp}, where the prime condensate forming candidates are \ce{SiO2}, \ce{MgSiO3}, Fe and \ce{Na2S}.
Although \ce{H2S} is currently not included in the chemical equilibrium model, it is likely to be a significant outgassed sulfur bearing species in the \ce{H2}-rich sub-Neptune environment, we therefore assume that there is sufficient \ce{H2S} in the atmosphere to efficiently form \ce{Na2S} once Na becomes saturated.
We assume the prime nucleation species is \ce{SiO2}, which is homogeneously nucleated assuming an efficient reaction process with available \ce{H2O} in the atmosphere.
We also considered homogeneous nucleation and growth of SiO directly using the vapour pressure and nucleation rate  expressions presented in \citet{Gail2013}, but find extrapolation from their considered temperature range (1275-1525 K) to the required 3000 K at the magma ocean surface to produce unrealistic results compared to \ce{SiO2}.
We therefore do not consider SiO condensation in the atmosphere in this study.

The cloud microphysical model evolves the vapour and condensate mass mixing ratios, $q_{\rm v}$ [g g$^{-1}$] and $q_{\rm c}$ [g g$^{-1}$] respectively, for each species, as well as the total particle volume mixing ratio, $q_{0}$ [cm$^{3}$ cm$^{-3}$].
Following \citet{Lee2025} the total number density, $N_{\rm c}$ [cm$^{-3}$], of the cloud particles is given by
\begin{equation}
    N_{\rm c} = q_{0} n_{\rm a},
\end{equation}
where $n_{\rm a}$ [cm$^{-3}$] is the number density of the atmosphere.
The number weighted mean mass of the particles, $m_{\rm c}$ [g], is given by
\begin{equation}
    m_{\rm c} = \frac{q_{\rm c}\rho_{\rm a}}{N_{\rm c}},
\end{equation}
where $\rho_{\rm a}$ [g cm$^{-3}$] is the mass density of the atmosphere.
The representative particle size, $r_{\rm c}$ [cm], is given by
\begin{equation}
    r_{\rm c} = \sqrt[3]{\frac{4m_{\rm c}}{3\pi\rho_{\rm d}}},
\end{equation}
where $\rho_{\rm d}$ [g cm$^{-3}$] is the bulk density of the cloud material. 
In this case, $\rho_{\rm d}$ is given by the volume weighted contribution of each individual species bulk density that is condensed onto the particles. 
Another quantity used in the model is the saturation mass mixing ratio, $q_{\rm s}$ [g g$^{-1}$], which is the local mass mixing ratio of the condensable vapour required to reach saturation in the atmosphere.

\section{Mineral clouds above magma oceans}
\label{sec:results}

\begin{figure*}
    \centering
    \includegraphics[width=0.49\linewidth]{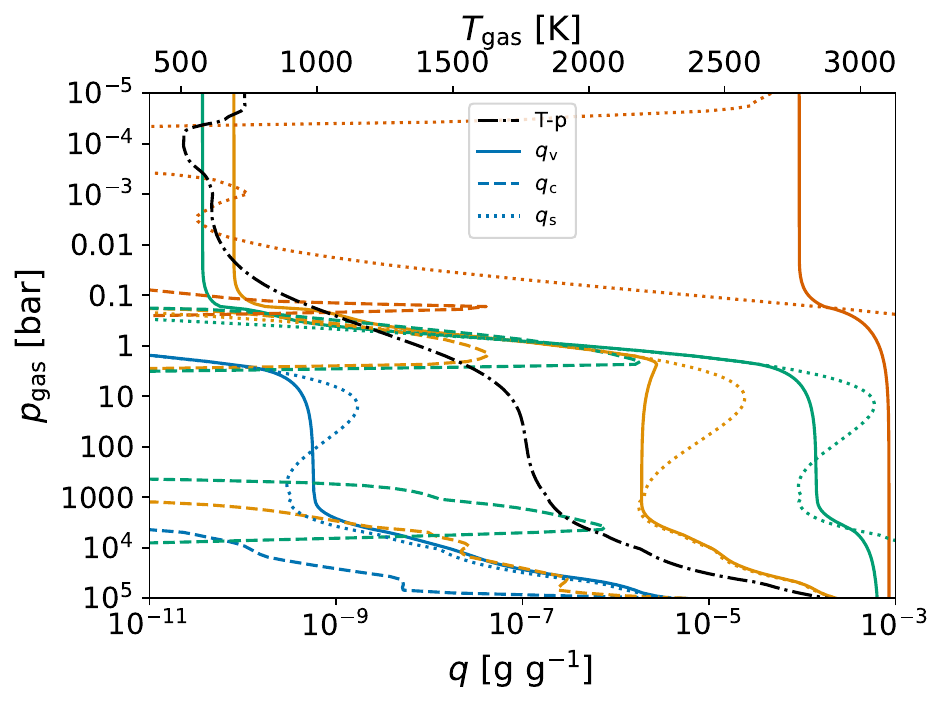}
    \includegraphics[width=0.49\linewidth]{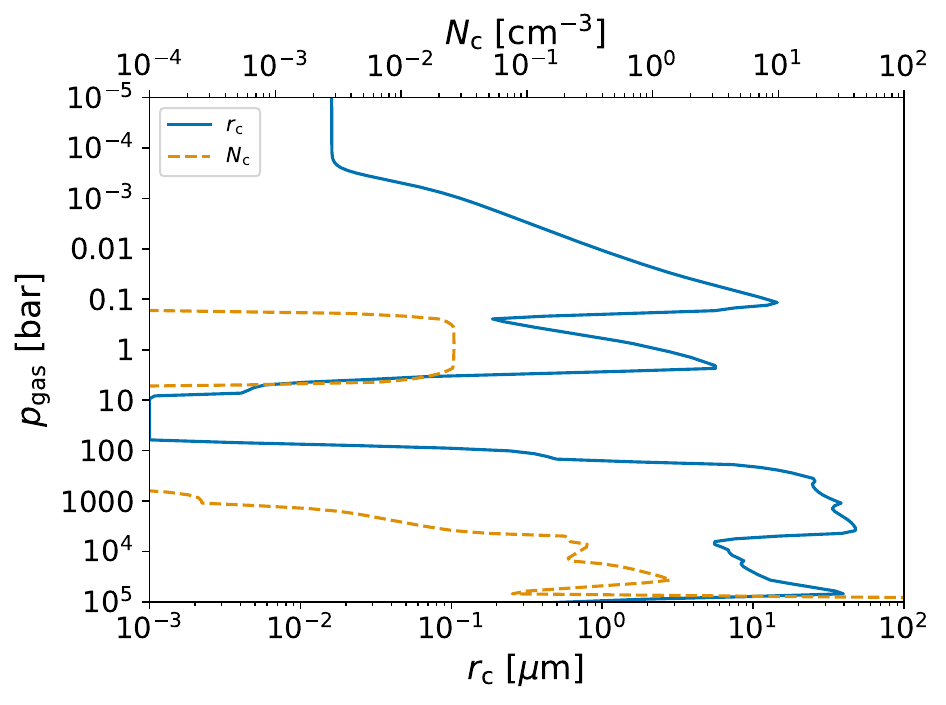}
    \includegraphics[width=0.49\linewidth]{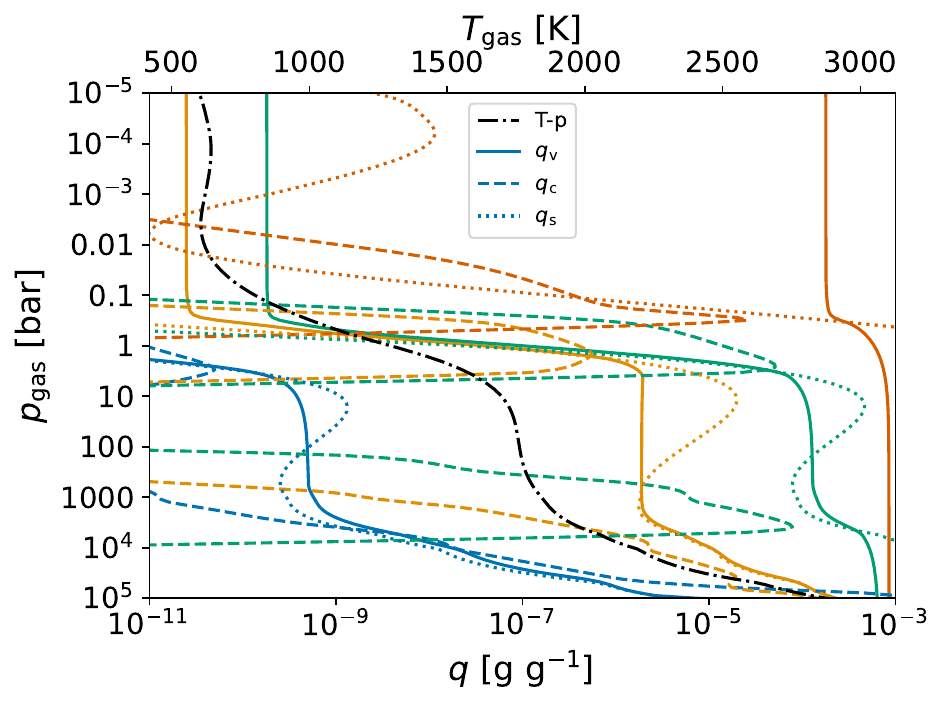}
    \includegraphics[width=0.49\linewidth]{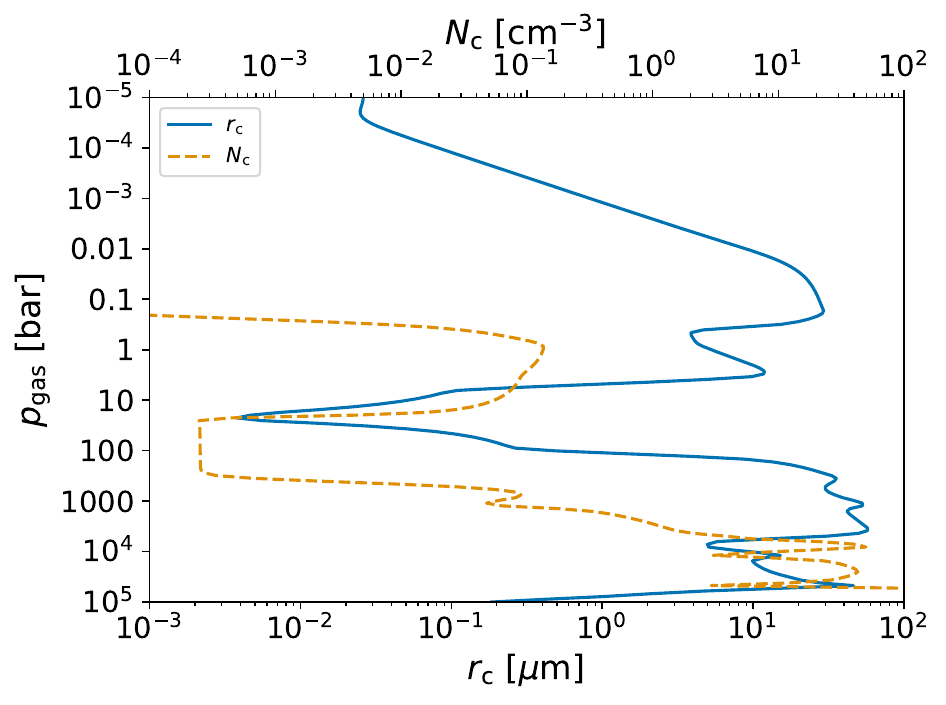}
    \includegraphics[width=0.49\linewidth]{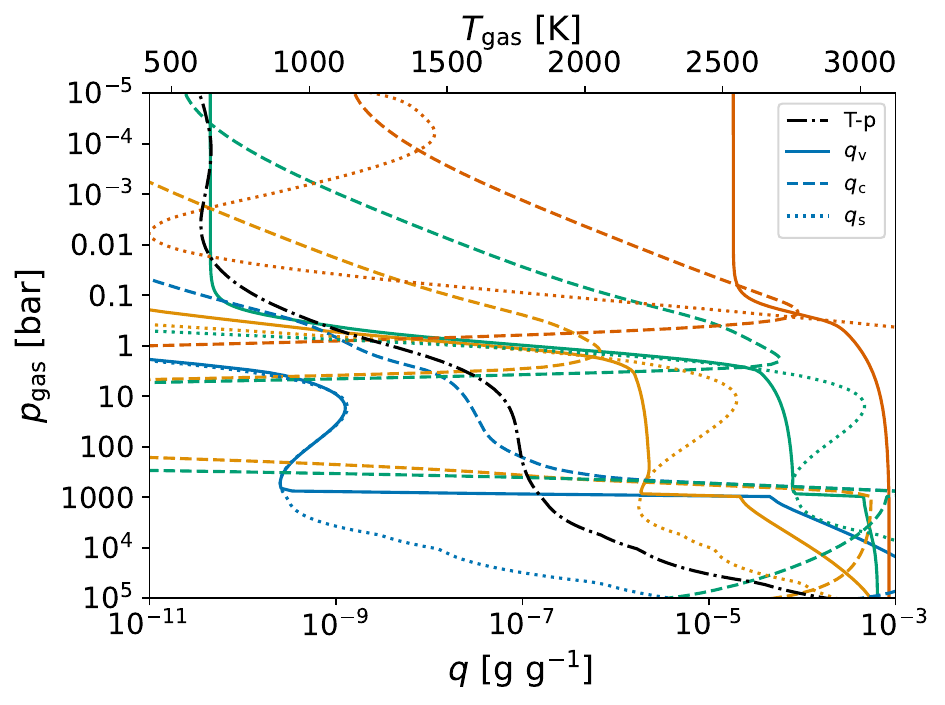}
    \includegraphics[width=0.49\linewidth]{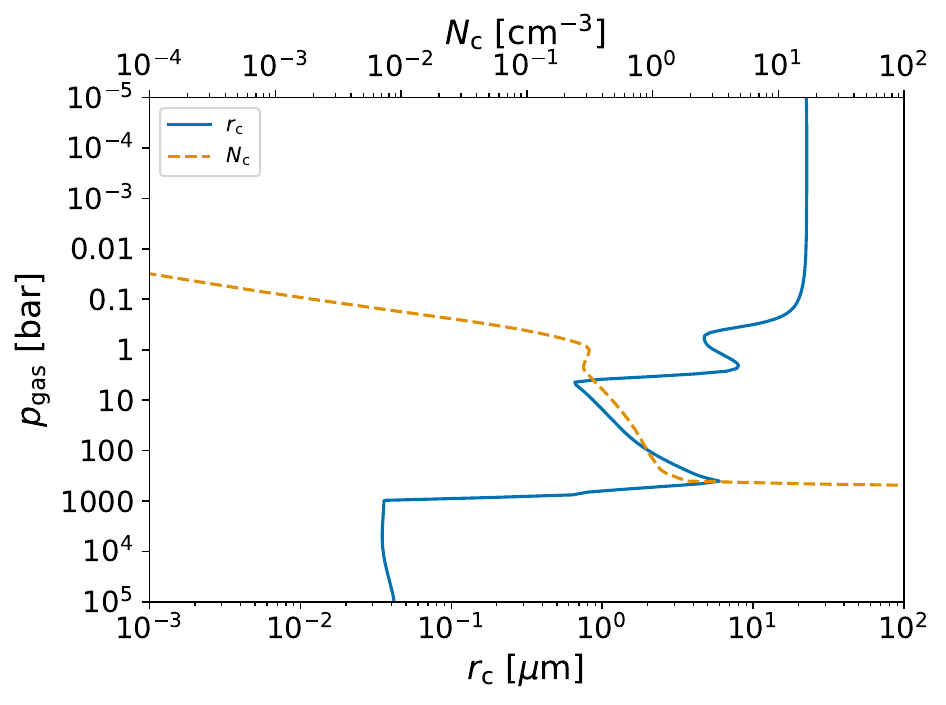}    
    \caption{Vertical cloud structures of the three mixing cases: $K_{\rm zz}$ = 10$^{4}$ cm$^{2}$ s$^{-1}$ (top row), $K_{\rm zz}$ = 10$^{7}$ cm$^{2}$ s$^{-1}$ (middle row) and vertically variable $K_{\rm zz}$ (bottom row).
    The left column shows the condensable vapor mass mixing ratio, $q_{\rm v}$ [g g$^{-1}$] (solid lines), condensate mass mixing ratio, $q_{\rm c}$ [g g$^{-1}$] (dashed lines), saturation mass mixing ratio, $q_{\rm s}$ [g g$^{-1}$] (dotted lines), and T-p profile (black dash-dot line).
    The colored lines denote the condensates \ce{SiO2} (blue), Fe (orange), \ce{MgSiO3} (green), and \ce{Na2S} (red).
    The right column shows the representative particle radius, $r_{\rm c}$ [$\mu$m] (blue solid line), and total particle number density, $N_{\rm c}$ [cm$^{-3}$] (orange dashed line) of the cloud particles.}
    \label{fig:1D_res}
\end{figure*}

\begin{figure*}
    \centering
    \includegraphics[width=0.49\linewidth]{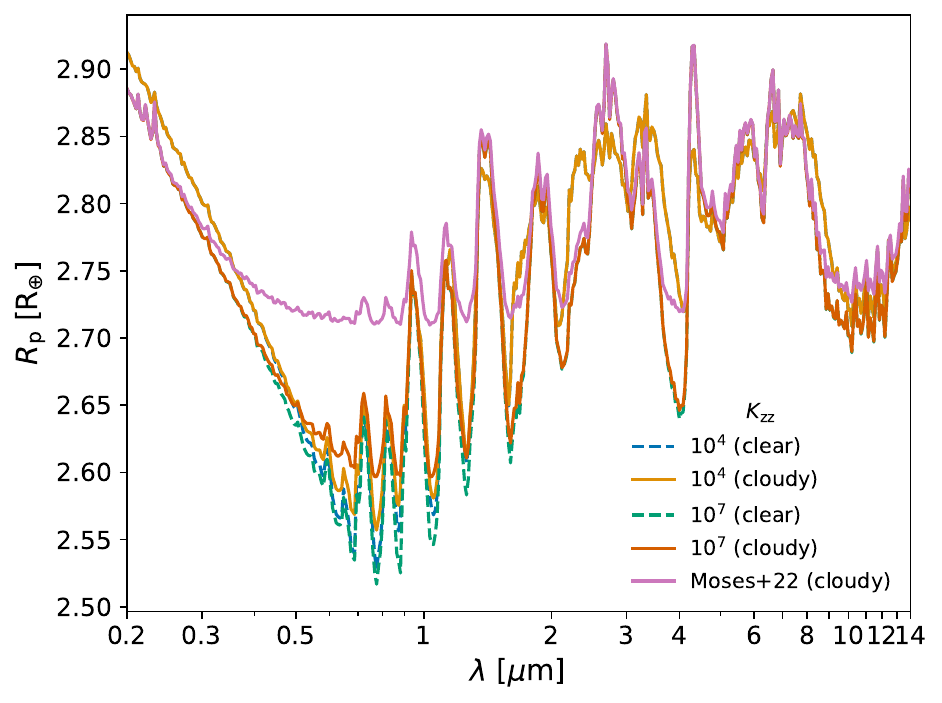}
    \includegraphics[width=0.49\linewidth]{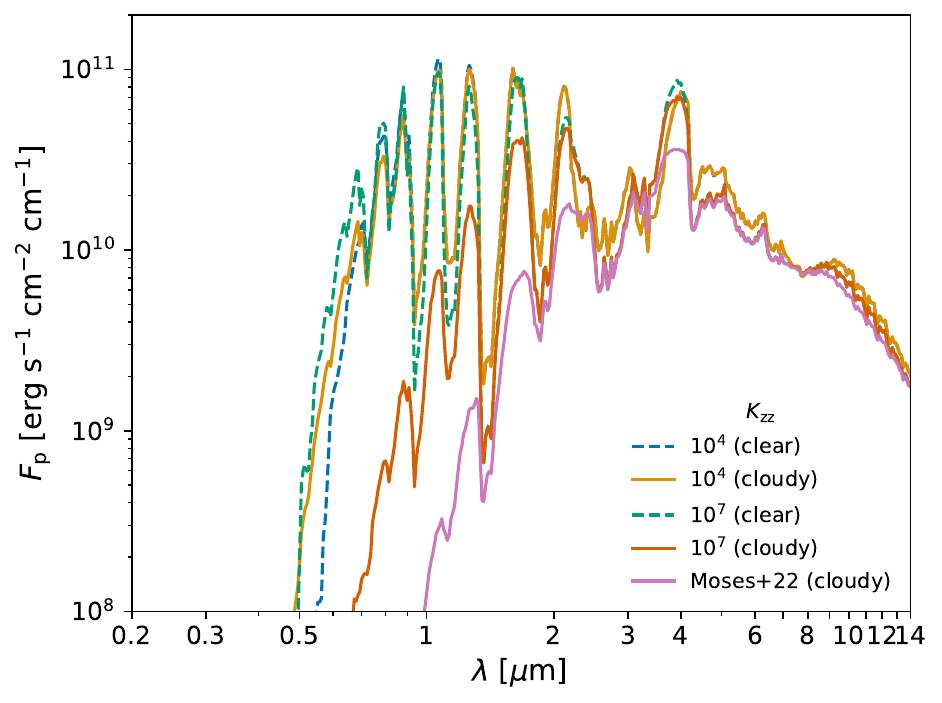}
    \includegraphics[width=0.49\linewidth]{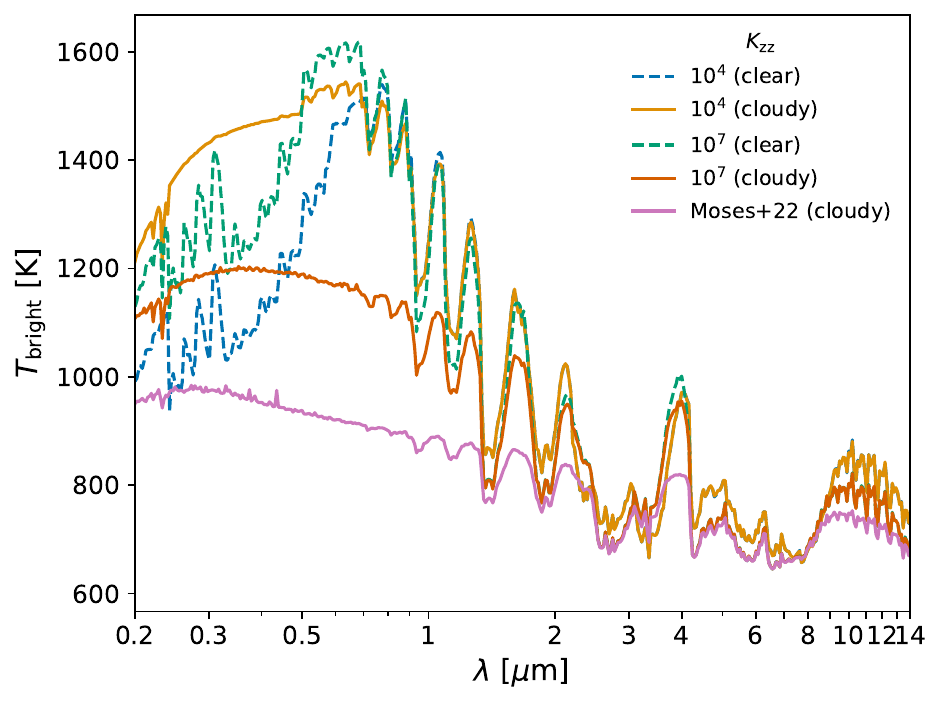}
    \includegraphics[width=0.49\linewidth]{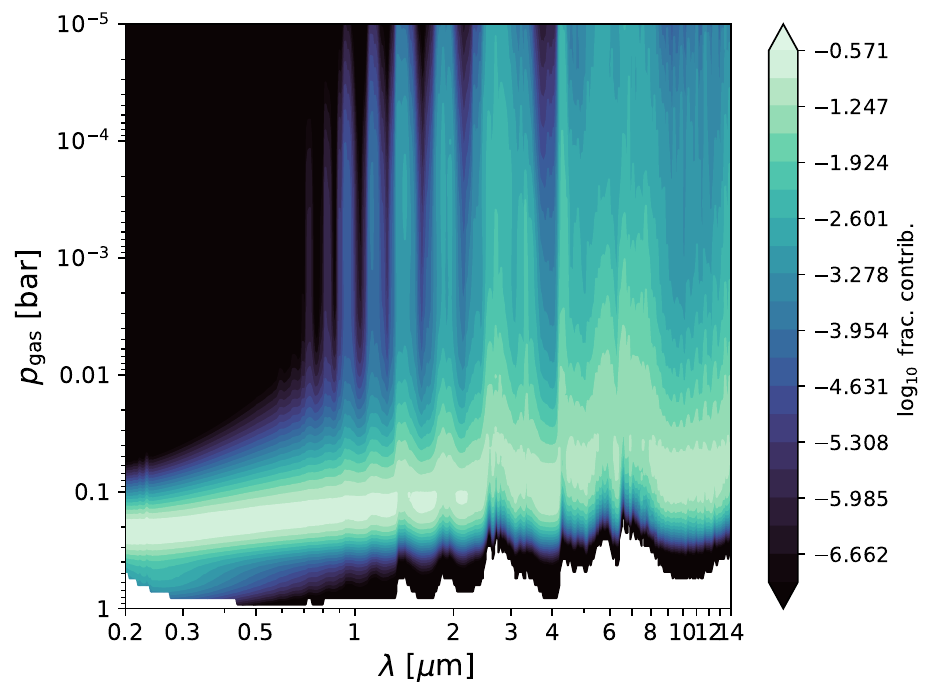}
    \caption{Top left: transmission spectra for all mixing cases with and without cloud opacity. 
    Top right: planetary emission spectra for all mixing cases with and without cloud opacity.
    Bottom left: brightness temperature, $T_{\rm bright}$ [K], of the planetary emission spectrum.
    Bottom right: contribution function of the emission spectra for the variable $K_{\rm zz}$ case.}
    \label{fig:spec}
\end{figure*}

We explore three vertical mixing rate cases, taking the T-p profile, atmospheric chemical and magma surface outgassing results from \citet{Werlen2025} as input to the cloud microphysical model.
Following \citet{Werlen2025}, one case considers a weak vertical mixing of $K_{\rm zz}$ = 10$^{4}$ cm$^{2}$ s$^{-1}$, while the other considers a moderate mixing rate of $K_{\rm zz}$ = 10$^{7}$ cm$^{2}$ s$^{-1}$.
We include a third case with a vertically varying $K_{\rm zz}$ profile, assuming a mixing profile for gaseous sub-Neptunes following Eq. (1) in \citet{Moses2022}, in which our profile parameters are $H_{\rm 1 mbar}$ $\approx$ 250 km and $T_{\rm eff}$ $\approx$ 885 K.
For this case we use the T-p profile and chemical composition from the $K_{\rm zz}$ = 10$^{7}$ cm$^{2}$ s$^{-1}$ case in \citet{Werlen2025}.
This produces a similar $K_{\rm zz}$ profile to the $T_{\rm eff}$ = 900 K case from \citet{Moses2022}.
We evolve each model until all layers show relative convergence of $<$ 10$^{-6}$ between each timestep, similar to the convergence criteria used for the VULCAN model \citep{Tsai2017}.

In Figure \ref{fig:1D_res}, we show the vertical cloud structure for our three vertical mixing rate cases.
Our results show that strong cloud formation and precipitation occurs directly above the magma-atmosphere boundary upward to a pressure of around 1000 bar, which is to be expected since vapour near the magma boundary is already at phase-equilibrium with the magma surface.
However, in all three cases we find the condensable vapour fraction follows closely the saturation vapour pressure profiles to the upper atmosphere, where it may condense again once saturated.
In the constant $K_{\rm zz}$ profile cases, this leads to two main cloud formation regions in the atmosphere, one in the deep atmosphere near the magma surface up to a pressure of $\sim$~100~bar and one in the upper atmosphere at pressures lower than $\sim$~10~bar.
In the constant weak mixing case, large particles of $\gtrsim$~10~$\mu$m are produced in the deep atmosphere cloud layer, while particles ranging from $\sim$~1-10~$\mu$m are present in the upper cloud layer.
Two discrete nucleation zones are present that contribute to the overall number density of the clouds, one just above the magma surface and one at around 1 bar, where \ce{SiO2} becomes saturated again.
In the constant moderate mixing case, a similar overall profile is produced, especially the deep cloud properties near the magma surface.
However, the upper cloud deck contains more number density and cloud particle sizes due to the stronger mixing enabling larger particles to remain lofted, as well as more rapid replenishment of vapour to the upper regions.
In the \citet{Moses2022} variable mixing profile case, the strong, assumed, convectively driven mixing in the deep atmosphere leads to very different deep atmosphere cloud structure with numerous amounts of particles of sizes $\sim$ 0.1 $\mu$m, suggesting very efficient nucleation of \ce{SiO2} in this case compared to the previous cases.
This case exhibits three distinct cloudy regions, the deep region described above, the intermediate region of weak mixing at around $\sim$ 1000-10 bar which prevents further lofting of material to the upper atmosphere, and a well mixed upper atmosphere where strong mixing returns.
In this case, there is therefore a much stronger connection to the condensate profile in the deeper regions to the upper atmosphere, rather than a vapour mixing driven effect in the weaker mixing cases.

In the weak and moderate mixing cases, efficient condensation of \ce{Na2S} in the upper atmospheric layers ($p_{\rm gas}$ $<$ 0.1 bar) is prevented due to a lack of cloud condensation surfaces, exhibited by the strong drop off in cloud number density below 0.1 bar, while \ce{Na2S} remains saturated.
This persists in the variable mixing case, but due to the increased mixing, larger particles are able to remain present all the way up to the upper boundary layers, though reduced in number density.

We produce synthetic transmission and emission spectra of our results using gCMCRT \citep{Lee2022}, with and without cloud opacity included.
The optical constants used for each cloud species are listed in Table \ref{tab:cond_sp}, the optical constants for species are mixed weighted by their relative bulk volume mixing ratio using effective medium theory with the Bruggeman method \citep[e.g.][]{Kiefer2024}.
We use the LX-MIE Mie theory code \citep{Kitzmann2018} to find the extinction efficiency, single scattering albedo and asymmetry factors of the cloud particles in each atmospheric layer.
We include opacity from the gas phase species \ce{CH4} \citep{Yurchenko2024}, \ce{CO} \citep{Li2015}, \ce{CO2} \citep{Yurchenko2020}, \ce{C2H2} \citep{Chubb2020}, \ce{H2O} \citep{Polyansky2018}, and \ce{OH} \citep{Mitev2025}, where the accompanying citation to each species is the line-list used to generate the input opacities to gCMCRT.
We include CIA opacity from \ce{H2}-\ce{H2} and \ce{H2}-He \citep{Karman2019} and Rayleigh scattering opacity from \ce{H2}, He and H.
For the multiple-scattering calculations, the Rayleigh phase function is sampled when gCMCRT detects a scattering event off the gas component and the Draine phase function \citep{Draine2003} is used for scattering off the cloud component.

Figure \ref{fig:spec} shows the transmission and emission spectra of our output. 
For the transmission spectra, in the weak and moderate mixing cases, the additional cloud opacity mutes features in the near-infrared by providing a baseline opacity that cuts off the opacity window of \ce{H2O} and \ce{CH4} features in the atmosphere.
The spectrum of the variable $K_{\rm zz}$ case is more affected by the cloud opacity, with a significant flattening of the spectrum across the optical and near-infrared wavelength regime, which greatly mutes the spectral features present in this wavelength range.
The mid-infrared is not greatly affected from the cloud structure in all cases.

Figure \ref{fig:spec} also shows the emission spectra for each case.
The weak mixing case has a limited effect on the emission spectra, while the moderate and variable mixing profiles produce redder spectra with a reduced flux compared to the cloud free spectrum. 
This can also be seen in the brightness temperature and contribution function plots in Figure \ref{fig:spec}, where it clear that the cloud opacity pushes the cooling regions to higher in the atmosphere.
The significant cloud opacity in the variable $K_{\rm zz}$ case leads to a contribution function that only reaches to pressures of $\sim$ 0.2 bar in the optical/near-IR wavelength regime.

\section{Discussion \& conclusions}
\label{sec:discon}

In this study, we apply a microphysical mineral cloud formation code to a sub-Neptune atmosphere for species outgassed from a deep surface magma ocean.
A significant uncertainty in the cloud microphysical model is the nucleation rate of seed particles, here chosen to be \ce{SiO2}, and the applicability of homogeneous classical nucleation theory applied in this study.
Should the nucleation rate be much larger than that modeled here, it is likely that more, smaller particles are produced in the atmosphere rather than the larger particles produced here.
A detection of small ($<$ 0.1 $\mu$m) silicate bearing clouds in cool sub-Neptunes could therefore be a potential indiction of an underlying magma ocean along with moderate/strong convective mixing present in the atmosphere, directly linking deep magma ocean processes to the observable atmospheric layers.

Our simple study does not include radiative-feedback from cloud opacity onto the atmosphere, magma ocean surface and interior structure, however, our results suggest this may be an important consideration given the strong effect on the emission spectrum from the cloud opacity. 
For the upper atmosphere, we suggest a similar temperature-pressure pattern to brown dwarf atmospheres, with heating occurring near cloud formation regions altering the T-p structure significantly \citep[e.g.][]{Morley2012, Morley2024}. 
This has critical implications also for the long-term cooling of sub-Neptunes \citep[e.g.][]{Lopez2014, ginzburg_super-earth_2016} and may further prolong the magma ocean states even for temperate sub-Neptunes.
In the deep atmosphere, it is less clear what the impact of the additional cloud opacity may be.
Due to very high pressures, \ce{H2}-{H2} CIA may still be the dominant opacity source, making the additional cloud opacity not important in setting the radiative gradient.
Other effects may include time-dependent cloud formation and feedback triggering deep convective motions in the atmosphere, these may induce dynamical vortex feature formation in the upper atmosphere, primarily dependent on the rotation rate regime of the planet, similar to those seen in brown dwarf GCM simulations \citep[e.g.][]{Lee2024}.

Our transmission spectra results suggest mineral clouds can significantly mute the spectral features, mimicking the effect of increased molecular weight of the atmosphere.
Characterizing these sub-Neptune atmospheres may therefore also suffer from the well known molecular weight - patchy cloud degeneracy similar to hot Jupiter exoplanets \citep{Line2016}.
This may be exacerbated if atmospheric dynamical processes lead to east-west terminator differences \citep[e.g.][]{Espinoza2024}, compared to our 1D global averaged profiles used in this study.

In this study, we do not consider the impact of the formation of photochemically produced haze particles on the cloud structure.
We suggest that settling haze particles from the low pressures regions in the atmosphere could interact with the excess mineral vapor in the atmosphere, providing additional condensation sites for haze-cloud hybrid particles to form, similar to that seen in Solar System gas giants.
This is likely to reduce the mean particle size of the particles in the upper atmosphere than those produced here, but the overall effect would depend significantly on the production rate of haze particles in the upper atmosphere \citep[e.g.][]{Steinrueck2023}.
Given the volatiles outgassed by the magma ocean, these haze particles may be composed of complex hydrocarbon chains \citep[e.g.][]{Moran2020}.
In addition, should enough \ce{H2S} be available in the atmosphere, \citet{Zahnle2009} suggest the photochemical formation of sulfur gases such as \ce{S2} in hot Jupiters. 
For similar to atmospheric conditions used in this paper, \citet{Gao2017} suggest this sulfur material may go on to form \ce{S2} and \ce{S8} complex hazes, which affect the albedo of the planet.
Furthermore, settling haze particles would increase muting of the transmission spectral features, which, combined with the cloud particles, would further mimic a high molecular weight atmosphere.

Recent studies have suggested that Silane (\ce{SiH4}), may be the dominant Si bearing outgassed species from the magma surface, rather than SiO in a reducing environment \citep{misener_atmospheres_2023,Ito2025}.
We suggest that this scenario would not alter substantially the main conclusions of the study, as the main reaction to form \ce{SiO2} clouds would change from SiO + \ce{H2O} $\rightarrow$ \ce{SiO2}[s] + \ce{H2} to \ce{SiH4} + 2\ce{H2O} $\rightarrow$ \ce{SiO2}[s] + 4\ce{H2}, with a similar change in the formation of \ce{MgSiO3} particles.
The overall efficiency of the reaction would reduce due to the requirement for more \ce{H2O} to react with \ce{SiH4}, but the overall end state cloud structures are likely to remain similar.

However, the presence of \ce{SiH4} instead of SiO may change the nucleation pathway and efficiency of seed particle formation if the further addition of a \ce{H2O} molecule alters the nucleation rate of \ce{SiO2} substantially.
This would have a significant effect on the cloud structure, especially number density and particle size, as our results suggest that the nucleation has a large effect on the final cloud profile.
Alternative nucleation species candidates that are able to form the initial seed particles would have to be considered, such as nucleation of \ce{MgSiO3} directly \citep{Goumans2012}.

\citet{Lupu2014} investigate the atmospheric composition resulting from a giant impact event that melts the surface of a terrestrial planet.
Their models suggest that the resulting atmosphere is similar in temperature and composition to that used in this study.
Given the capacity to easily form mineral clouds from the outgassed materials, it is plausible that similar mineral clouds would be able to be form in the atmosphere after such an impact event from the resulting surface magma ocean.
In addition, they suggest photochemical products such as \ce{SO2} would be an important feature of the atmospheric consequences of such a collision.
However, we note that \citet{Lupu2014} considered a \ce{N2} dominated background gas, rather than the \ce{H2} considered here.

In conclusion, our study has found that mineral clouds can easily form in the atmosphere above sub-Neptune magma surface, due to the outgassed refractory material mixing upward to cooler regions of the atmosphere and condensing.
Clouds composed of \ce{SiO2}, \ce{Fe}, \ce{MgSiO3} mixtures form near the magma surface to pressures of $\sim$ 1000 bar, while a secondary cloud layer forms at around $\sim$ 0.1 bar, primarily composed of \ce{Na2S}.
With moderate to strong mixing in the atmosphere, these cloud particles can greatly affect the transmission spectra, shallowing the spectral features at near-infrared wavelengths. 
We find the large particle sizes produced by the model do not show strong signs of cloud absorption features beyond $\sim$2 $\mu$m.
In emission spectra, these clouds can also redden and reduce the overall flux from escaping the atmosphere.
This suggests clouds may play an important part in the complex atmosphere-surface-interior system in these sub-Neptunes, altering the atmospheric T-p structure which may feed back into the surface conditions as well as altering their observed properties significantly.
Furthermore, clouds affect the long-term evolution of sub-Neptunes. Gigayear timescales for magma-ocean cooling are predicted for the majority of observed sub-Neptunes due to the insulation of a thick atmosphere even without clouds. 
The presence of high altitude clouds may further reduce the cooling efficiency of sub-Neptunes and thereby extend the possibility to have magma oceans on temperate sub-Neptune. 

\begin{acknowledgments}
E.K.H. Lee acknowledges support from the CSH through the Bernoulli Fellowship. 
C.D acknowledges support from the Swiss National Science Foundation under grant TMSGI2\_211313. 
This work has been carried out within the framework of the NCCR PlanetS supported by the Swiss National Science Foundation under grant 51NF40\_205606.
\end{acknowledgments}

\begin{contribution}
E.K.H. Lee performed the cloud microphysical modeling and production of spectra.
A. Werlen and C. Dorn provided the temperature-pressure and chemical profile input data used for the microphysical model.
\end{contribution}

\software{mini-cloud\footnote{\url{https://github.com/ELeeAstro/mini_cloud}} \citep{Lee2025}, gCMCRT\footnote{\url{https://github.com/ELeeAstro/gCMCRT}} \citep{Lee2022}}

\bibliography{bib_aaron,bib}
\bibliographystyle{aasjournalv7}

\end{document}